\documentclass[11pt,epsfig]{article}
\usepackage{epsfig}
\usepackage{amsfonts}
\usepackage{amsmath}
\usepackage{eucal}
\usepackage{amssymb}

\newcommand{\newsection}{ \setcounter{equation}{0} \section}
\def\appendix#1{
  \addtocounter{section}{1}
  \setcounter{equation}{0}
  \renewcommand{\thesection}{\Alph{section}}
  \section*{Appendix \thesection\protect\indent #1}
  \addcontentsline{toc}{section}{Appendix \thesection\ \ \ #1}
  }
\newcommand{\beq}{\begin{equation}}
\newcommand{\eeq}{\end{equation}}
\newcommand{\eqn}{\begin{eqnarray}}
\newcommand{\feqn}{\end{eqnarray}}
\newtheorem{note}{Remark}[section]
\newtheorem{lemma}{Lemma}[section]
\newtheorem{corollary}{Corollary}[section]
\newtheorem{proposition}{Proposition}[section]

\newenvironment{teo}[1]{\begin{theorem}#1}{\end{theorem}}

\newenvironment{lem}[1]{\begin{lemma}#1}{\end{lemma}}
\newenvironment{cor}{\begin{corollary}}{\end{corollary}}
\newenvironment{prop}{\begin{proposition}}{\end{proposition}} 

\newcommand{\eps}{\varepsilon}

\newcommand{\Bb}{\mbox{\rm\bf B} }
\newcommand{\Db}{\mbox{\rm\bf D} }
\newcommand{\Eb}{\mbox{\rm\bf E} }
\newcommand{\Hb}{\mbox{\rm\bf H} }

\newcommand{\epsb}{\mbox{\boldmath $\eps$} }
\newcommand{\betb}{\mbox{\boldmath $\beta$} } 

\newcommand{\hA}{\hat{A}}
\newcommand{\hf}{\hat{f}}
\newcommand{\hF}{\hat{F}}
\newcommand{\hL}{\hat{{\cal L}}}
\newcommand{\hg}{\hat{g}}

\newcommand{\ba}{\begin{array}}
\newcommand{\ea}{\end{array}}

\newtheorem{dfn}{Definition}[section]

\begin{document}
\topmargin 0pt
\oddsidemargin 5mm
\headheight 0pt
\headsep 0pt
\topskip 9mm
 
\addtolength{\baselineskip}{0.20\baselineskip}

\hfill IFUM-728/FT

\hfill October 2002

\begin{center}
 
\vspace{88pt}

{\Large \bf
Noncommutative Electrodynamics\\
}

\vspace{40pt}
\renewcommand{\thefootnote}{\fnsymbol{footnote}}
{\large \bf {G.~Berrino$^{1}$\footnote{berrino@vmimat.mat.unimi.it},
S.\ L.~Cacciatori$^{1,3}$\footnote{cacciatori@mi.infn.it},
A.~Celi$^{2,3}$\footnote{alessio.celi@mi.infn.it},
L.~Martucci$^{2,3}$\footnote{luca.martucci@mi.infn.it} and
A.~Vicini$^{2,3}$\footnote{alessandro.vicini@mi.infn.it}}}\\
\renewcommand{\thefootnote}{\arabic{footnote}}
\setcounter{footnote}{0}
\vskip 10mm
{\small
$^1$ Dipartimento di Matematica dell'Universit\`a di Milano,\\
Via Saldini 50, I-20133 Milano.\\

\vspace*{0.5cm}

$^2$ Dipartimento di Fisica dell'Universit\`a di Milano,\\
Via Celoria 16, I-20133 Milano.\\

\vspace*{0.5cm}

$^3$ INFN, Sezione di Milano,\\
Via Celoria 16,
I-20133 Milano.\\
}

\end{center}
  
\vspace{25pt}

\begin{center}

{\bf Abstract}

\end{center}

{\small 
In this paper we define a causal Lorentz covariant noncommutative (NC)
classical Electrodynamics.
We obtain an explicit realization of the NC theory
by solving perturbatively the Seiberg-Witten map. 
The action is polynomial in the field strenght $F$,
allowing to preserve both causality and Lorentz covariance.
The general structure of the Lagrangian is studied, to all orders in the
perturbative expansion in the NC parameter $\theta$.
We show that monochromatic plane waves are solutions of the equations
of motion to all orders.
An iterative method has been developed to solve the equations of motion and
has been applied to the study of the corrections to the superposition law and
to the Coulomb law.

\vspace{35pt}


\vfill
 
\newpage
 
\newsection{Introduction}

Suggestions on the possibility that Nature could allow for noncommuting spatial 
coordinates, came both from the past \cite{Sn} and more recently in the realm of
superstring theory studying low energy excitations of $D$-branes in a magnetic 
field \cite{SW}. \par
This has stimulated investigations on the noncommutative (NC) versions of
gauge field theories and on the behaviour of their quantized counterparts.
Among these, Maxwell theory is perhaps the easiest example and one where
a possible experimental test of this hypothesis could be realizable.
Nevertheless, two main problems arise when one tries to implement 
Electromagnetism in a noncommutative geometry: the loss of causality due 
to the appearence of derivative couplings in the Lagrangian and, more 
fundamentally, the violation of Lorentz invariance exhibited by plane wave 
solutions \cite{JK1}.
These problems have been discussed with a different approach in the
framework of NC QED \cite{carone,Mo}.
\par
In this paper we show that both these problems may be avoided if one allows a
nonzero "electrical" component into the tensor $\theta$ of the noncommutation
relations so including time as a NC coordinate. After application of the
Seiberg-Witten map \cite{SW} the theory is perturbative in $\theta$ and 
classical plane waves turn out to be exact solutions. They no longer
obey a superposition principle. Finally a sort of Electric-Magnetic 
duality comprehending $\theta$ and reminiscent of the known one in commutative
Maxwell theory, appears between the fields in the equations of motion 
reinforcing the interpretation of $\theta$ as a sort of background primordial
Electromagnetic field.
\par
In Section 2 we fix notations and conventions,
recall the definition of the S-W map \cite{SW} 
and show the explicit solution to second order in $\theta$.\par
In Section 3 we prove that the Lagrangian of the theory is polynomial to all
orders in the perturbative parameter so that causality is preserved.
The equations of motion are derived in Section 4 where evidence is also given
of the mentioned duality.\par
In Section 5 a general iterative method of solving the equations of motion
is outlined. \par\noindent
 After proving that plane waves are solutions, the method is
applied to the problem of plane wave superposition and to derive corrections
to the Coulomb law.\par
The paper ends with some comments on the results found and on possible
experimental settings aimed to directly measure noncommutativity.  

\newsection{The S-W map and second order expantions}

In the following, a hat over a classical symbol will indicate the 
same quantity in its NC version. In this fashion, coordinates of flat 
noncommutative Minkowsky spacetime will be assumed to satisfy:
\begin{equation}
[\hat{x}^\mu,\hat{x}^\nu]_*=i\,\theta^{\mu\nu}
\end{equation}
where $\theta^{\mu\nu}$ is a real skew tensor whose components are set as 
follows:
\begin{equation} \label{convention}
\left\{\begin{array}{rcl}
\theta^{0i} & = & \eps^i  \\
\theta^{ij} & = & \epsilon^{ijk}\,\beta_k 
\end{array}\right.
\end{equation}
Note that we do not impose $\eps^i= 0$. This means that time does not commute
with spatial coordinates and $\theta$ is a constant tensor field.
Besides $\theta$ we consider the usual Electromagnetic field whose NC action 
is given by:
\begin{equation}  \label{action}
\hat{S}=-\frac{1}{4}\int d^4x\,\hF^{\mu\nu}*\hF_{\mu\nu}=
-\frac{1}{4}\int d^4x\,\hF^{\mu\nu}\,\hF_{\mu\nu}
\end{equation}
The corresponding Lagrangian and field strenght are given by:
\begin{eqnarray}
\hL & = & -\frac{1}{4}\,\hF^{\mu\nu}\,\hF_{\mu\nu}  \label{Lagrangian} \\
\hF_{\mu\nu} & = & \partial_\mu\hA_\nu-\partial_\nu\hA
_\mu-i\left[\hA_\mu,\hA_\nu\right]_*  
\end{eqnarray}
Here the star-product ($*$) between NC quantities is defined as usual:
\begin{equation}
(\hf * \hg)(x):=e^{\frac i2\theta^{\mu\nu}\partial_\mu\partial'_\nu}\hf(x)
\hg(x')_{|x=x'}
\end{equation} 
Also, the following conventions will be used for Electromagnetic fields:
\begin{equation}
\begin{array}{rcl}
E^i & = & F^{0i} \\
B_k & = & \frac{1}{2}\,\epsilon_{ijk}\,F^{ij} 
\end{array}
\end{equation}
Now, according to Seiberg and Witten \cite{SW}, every NC gauge theory 
$\hA_\mu$ has a perturbative description in terms of the non commuting 
parameter $\theta$ and another commutative theory $A_\mu$ possessing 
the same degrees of freedom as the NC one. The relation between them is
established by means of the Seiberg-Witten map:
\begin{equation}  \label{SWmap}
\left\{\begin{array}{rcl}
\frac{\partial\hA_{\mu}}{\partial\,\theta^{\alpha\beta}} & = & -\frac{1}{8}\,
\{\hA_{\alpha},\partial_\beta\hA_{\mu}+\hF_{\beta\mu}\}_*-(\alpha
\leftrightarrow\beta) \\
\hA_{\mu|\theta=0} & = & A_{\mu}
\end{array}\right.
\end{equation}
Solving the above equations means determine each piece of the perturbative
expansions:
\begin{eqnarray}
\hA_{\mu} & = & A_{\mu}+\hA_{\mu}^{(1)}+\hA_{\mu}^{(2)}+\cdots \\
\hF_{\mu\nu} & = & F_{\mu\nu} + \hF_{\mu\nu}^{(1)} + \hF_{\mu\nu}^{(2)} 
+\cdots      \label{field}
\end{eqnarray}
relating at every order in $\theta$ the NC quantities with their respective
classical counterparts. As is well known \cite{Bi}, one obtains to first order:
\begin{equation}
\begin{array}{rcl}
\hA_{\mu}^{(1)} & = & -\frac{1}{2}\,\theta^{\alpha\beta}A_\alpha
(\partial_\beta A_\mu+F_{\beta\mu}) \\
\hF_{\mu\nu}^{(1)} & = & \theta^{\gamma\delta}\left(F_{\mu
\gamma}F_{\nu\delta}-A_\gamma\partial_\delta F_{\mu\nu}\right)
\end{array}
\end{equation}
Considering the second order corrections, we assume $\hA_{\mu}^{(2)}=
\frac{1}{2}\,\theta^{\alpha\beta}\theta^{\gamma\delta}\mathfrak{n}_
{\mu\alpha\beta\gamma\delta}$ and substitute the whole expansion of 
$\hA_{\mu}$ into (\ref{SWmap}). We realize that differentiating and then
evaluating at $\theta=0$, we end with a recursive relation between 
second order and first order corrections and their $\theta$-derivatives.
This leads to computation of the term $\mathfrak{n}$. After careful
rearrangements, the expression for the second order correction to 
$A_\mu$ is:
\begin{equation}
\hA_{\mu}^{(2)} = \frac{1}{2}\,\theta^{\alpha\beta}\theta^{\gamma\delta}
\left\{A_\gamma\partial_\delta A_\alpha\partial_\beta A_\mu + A_\gamma
F_{\delta\alpha}F_{\beta\mu} + A_\alpha A_\gamma\partial_\delta F_{\beta\mu} 
+\frac{1}{4}\,\partial_\mu
(A_\alpha A_\gamma\partial_\delta A_\beta)\right\} 
\end{equation}
Similarly, via the relation $\hF_{\mu\nu}^{(2)}=\partial_\mu\hA_{\nu}^{(2)}+
\theta^{\gamma\delta}\partial_\gamma\hA_{\mu}^{(1)}\,\partial_\delta 
A_\nu -(\mu\leftrightarrow\nu)$ one also computes:
\begin{equation}
\hF_{\mu\nu}^{(2)} = \theta^{\alpha\beta}\theta^{\gamma\delta}\,F_{\mu\gamma}
F_{\delta\alpha}F_{\beta\nu}-\theta^{\gamma\delta}\,A_\gamma \partial_\delta 
\hF_{\mu\nu}^{(1)}-\frac{1}{2}\,\theta^{\alpha\beta}\theta^{\gamma\delta}
A_\gamma\left(\partial_\alpha A_\delta + A_\alpha\partial
_\delta\right)\partial_\beta F_{\mu\nu}
\end{equation}

\newsection{The general structure}
We discuss some properties valid to all orders in $\theta$ 
of the perturbative action 
obtained by means of the S-W map.  
\begin{prop}
The Lagrangian $\hL$ corresponding to the action (\ref{action}) via the S-W 
map is a polynomial in $F$ only (that is: it does not contain derivatives 
of $F$); furthermore the terms $\hL^{(n)}$ of order $n$ in $\theta$ form a 
homogeneous polynomial of degree $n+2$ in $F$.
\end{prop}{}
{\em Proof :}\\
  From the S-W equation (\ref{SWmap}) we have:
\eqn
& & \frac {\delta \hF_{\mu \nu}}{\delta\,\theta^{\alpha \beta}} =
\frac 18 \,\partial_\nu
\left\{ \hat A_\alpha , \partial_\beta \hat A_\mu
+\hat F_{\mu \nu} \right\}_*
-(\mu \leftrightarrow \nu)
+\frac i8 \left[
\left\{ \hat A_\alpha , \partial_\beta \hat A_\mu
+\hat F_{\beta \mu} \right\}_* ,\hat A_\nu \right]_* \cr
& & +\frac i8 \left[ \hat A_\mu ,
\left\{ \hat A_\alpha , \partial_\beta \hat A_\nu
+\hat F_{\beta \nu} \right\}_* \right]_*
-(\alpha \leftrightarrow \beta) +\delta * \label{dimostra}
\feqn
Here $\delta *$ is supposed to include all the terms arising whenever the 
derivation acts on the $\theta s$ appearing in the $*$ of the $*$ - product; 
they always give rise to total derivatives in the Lagrangian density and so 
may be neglected.
As a consequence, performing an arbitrary number of derivations and then 
putting $\theta =0$ shows that commutators of the type present in 
(\ref{dimostra}) give vanishing contributions. \\
Then all significant contributions are seen to come from the term
$\frac 18 \,\partial_\nu
\left\{ \hat A_\alpha , \partial_\beta \hat A_\mu
+\hat F_{\mu \nu} \right\}_* $
which, evaluated at $0$ after $k$ derivations, produces an homogeneous 
polynomial of order $k+1$ in $A$ with $k+1$ derivatives (with respect to 
spacetime coordinates) equally distributed on each monomial. 
Finally, considering $\hat F^{\mu \nu} * \hat F_{\mu \nu}$ at order $n$ 
in $\theta$, by the same argument, one obtains an homogeneous polynomial
of order $n+2$ in $A$ with $n+2$ spacetime derivatives comparing in each
monomial. \\
Now, since the Lagrangian density (obtained from the S-W map) is certainly 
invariant under the usual $U(1)$ gauge transformations, every monomial
can be rearranged, modulo integration by parts, so as to depend only on
$F$ and possibly its derivatives. But being the number of derivatives 
exactly equal to the number of $As$ in every monomial, it follows that 
derivatives of $F$ cannot appear at all.     QED

\begin{cor}
The equations of motion of the $U(1)$ theory take the form:
\eqn
\partial_\nu \tilde F^{\mu \nu} =0
\feqn
where $\tilde F^{\mu \nu}$ is the sum of homogeneous polynomials of degree 
$n+1$ in $F$ and order $n$ in $\theta$ (i.e. written symbolically):
\eqn   \label{struct}
\tilde F =\sum_n \theta^n F^{n+1} 
\feqn
\end{cor}
As we will see this property helps to derive a recursive algorithm for 
their resolution.

The main consequence of the structure (\ref{struct}) evidenced above, is
that the equations of motion for the field strenght are of first order.
This seems to suggest that the theory is causal even though not
requiring time commutativity. In the literature, it is suspected that 
causality does not survive Noncommutativity \cite{Bz}. In our model though, 
after undertaking the SW map, the action has been manipulated and integrated 
by parts to render all terms explicitly gauge invariant.
As a by-product, all higher order time derivatives have disappeared.
In effect, this task is equivalent to add boundary terms to the
Lagrangian: exactly those capable of giving causal consistency to the theory.\par
Probably this should be the right procedure to follow generally.
Furthermore, the fact that preserving causality is no more consequence of
imposing zero temporal components in $\theta$, allows to require that it
can transform like a tensor in respect to the Lorentz group. It descends
that Lorentz covariance is also preserved.

\newsection{Equations of motion up to second order}

Let us expand also the NC Lagrangian density (\ref{Lagrangian}) into pieces 
of increasing order in $\theta$:
\begin{equation}
\hL={\cal L}+{\hL}^{(1)}+{\hL}^{(2)}+\cdots 
\end{equation}
The first term here coincides with the classical Maxwell Lagrangian while
the other terms are its various corrections. More precisely:
\begin{eqnarray*}
{\cal L}  & = & -\frac{1}{4}\,F^{\mu\nu}\,F_{\mu\nu} \\
\hL^{(1)} & = & -\frac{1}{2}\,\hF^{\mu\nu (1)}\,F_{\mu\nu} \\
\hL^{(2)} & = & -\frac{1}{4}\left\{\hF^{\mu\nu (1)}\hF_{\mu\nu}^{(1)} + 
                      2 F^{\mu\nu}\,\hF_{\mu\nu}^{(2)}\right\}
\end{eqnarray*}
Recall \cite{Kv} that up to first order in $\theta$, the NC Lagrangian 
has the following form:
\begin{equation}
\hL = -\frac{1}{4}\,\left[\left(1-\frac{1}{2}~\theta^{\alpha\beta} 
F_{\alpha\beta}\right)F^2+2\,\theta^{\alpha\beta}F^{\mu\nu}F_{\mu\alpha}
F_{\nu\beta}\right]
\end{equation}
or, upon substitution according to our conventions (\ref{convention}) we have:
\begin{equation}
\hL = \frac{1}{2}\,(1+\betb\cdot\Bb-\epsb\cdot\Eb)(\Eb^2-\Bb^2)
-(\betb\cdot\Eb+\epsb\cdot\Bb)(\Eb\cdot\Bb)
\end{equation}
Next, looking for the second order term, one finds with a little effort:
\begin{eqnarray}
\hL^{(2)} & = & -\frac{1}{4}\,\theta^{\alpha\beta}\theta^{\gamma\delta}
\left\{F^\mu_{~\alpha}F^\nu_{~\beta}F_{\mu\gamma}F_{\nu\delta} + 2\,F^{\mu\nu}
F_{\mu\gamma}F_{\beta\nu}F_{\delta\alpha}+F^{\mu\nu}F_{\mu\alpha}F_{\nu\beta}
F_{\delta\gamma}\,+\right. \nonumber \\
&  & +\,\frac{1}{8}F_{\beta\alpha}F_{\delta\gamma}F^2+\frac{1}{4}
F_{\beta\gamma}F_{\alpha\delta}F^2\left.\right\}
\end{eqnarray}
Here again, after substitution and accurate computation, you get:
\begin{eqnarray}
\hL^{(2)} & = & (\epsb\cdot\Eb-\betb\cdot\Bb)(\betb\cdot\Eb+\epsb\cdot\Bb)
(\Eb\cdot\Bb)+\frac{1}{2}\,\left[(\epsb\cdot\Eb-\betb\cdot\Bb)^2(\Eb^2-\Bb^2) 
\,+ \right. \nonumber \\
& & +\,(\epsb\cdot\betb)(\Eb^2-\Bb^2)(\Eb\cdot\Bb)-(\Eb\cdot\Bb)^2(\epsb^2
-\betb^2) \left. \right]
\end{eqnarray}
As already remarked, the second variation of $\hL$ yields the usual equations
of motion:
\begin{equation} \label{motion}
\partial_\nu\tilde{F}^{\mu\nu}=0
\end{equation}
and eq.(\ref{struct}) leads us to write $\tilde{F}=F+\tilde{F}^{(1)}+\tilde{F}
^{(2)}+\cdots$ where $\tilde{F}^{(n)}\equiv\,\theta^n\,F^{n+1}$. \par
It is now tempting to regard each piece like this as a correction to the 
classical field strenght $F$ due to NC geometry. This is more properly done 
here than on the expansion (\ref{field}) because we are referring to the
equations of motion. Furthermore the interesting thing \cite{JK1} is that 
denoting the content of the NC field $\tilde{F}$ with an Electric displacement 
and Magnetic induction $(\Db,\Hb)$ and restating the above expansion as:
\begin{eqnarray*}
\Db & = & \Eb+\Db^{(1)}+\Db^{(2)}+\cdots \\
\Hb & = & \Bb+\Hb^{(1)}+\Hb^{(2)}+\cdots
\end{eqnarray*} 
where the classical fields $(\Eb,\Bb)$ in $F$ are recaptured as their zeroth
order corrections $(\Db^{(0)},\Hb^{(0)})$, then the equations of motion take 
the usual Maxwell form:
\begin{equation}
\begin{array}{rcl}
{\displaystyle \frac{\partial\,\Bb}{\partial\,t}\,+\,\nabla\times\Eb} & = & 0 \\
{\displaystyle \nabla\cdot\Bb } & = & 0 
\end{array}
\end{equation}
\begin{equation}
\begin{array}{rcl}
{\displaystyle \frac{\partial\,\Db}{\partial\,t}\,-\,\nabla\times\Hb} & = & 0 \\
{\displaystyle \nabla\cdot\Db } & = & 0
\end{array}
\end{equation}
Note that the first two are simply the Bianchi identities; 
the other two really describe the behaviour 
of NC Electromagnetism in empty space. Working with the first order correction
to $F$ which is:
\begin{equation} \label{first}
\tilde{F}^{\mu\nu(1)}=-\frac 12 \,(\theta F)F^{\mu\nu}-\frac 14 \,
\theta^{\mu\nu}F^2+\theta_{\alpha\beta}F^{\mu\alpha}F^{\nu\beta}+
(\theta^{\mu\beta}F^{\alpha\nu}-\theta^{\nu\beta}F^{\alpha\mu})
F_{\alpha\beta}
\end{equation}
we obtain for the NC fields the approximated expressions:
\begin{equation} \label{NC}
\begin{array}{rcl}
\Db & = & (1+\betb\cdot\Bb-\epsb\cdot\Eb)\,\Eb-(\betb\cdot\Eb+\epsb\cdot\Bb)\,\Bb
-\frac{1}{2}\,(\Eb^2-\Bb^2)\,\epsb-(\Eb\cdot\Bb)\,\betb \\
\\
\Hb & = & (1+\betb\cdot\Bb-\epsb\cdot\Eb)\,\Bb+(\betb\cdot\Eb+\epsb\cdot\Bb)\,\Eb
-\frac{1}{2}\,(\Eb^2-\Bb^2)\,\betb+(\Eb\cdot\Bb)\,\epsb
\end{array}
\end{equation}
Here the NC tensor $\theta$ has been assumed to represent a couple of fields
(\epsb,\,\betb) in agreement with the conventions (\ref{convention}). 
The second order correction to $F$ reads explicitly:
\begin{eqnarray}
\tilde{F}^{\mu\nu(2)} & = & \frac 14 \,\theta^{\alpha\beta}F^{\gamma}_{~\delta}
F_{\gamma\beta}(\theta^{\nu\delta}F^{\mu}_{~\alpha}-\theta^{\mu\delta}F^{\nu}
_{~\alpha}) \nonumber \\
& + & \frac 14 \left\{\theta^{\alpha\beta}\theta^{\gamma\delta}F^{\mu}_{~\gamma}
F^{\nu}_{~\beta}F_{\alpha\delta}+\theta^{\alpha\beta}F_{\beta\gamma}
F_{\delta\alpha}(\theta^{\nu\delta}F^{\mu\gamma}-\theta^{\mu\delta}F^{\nu\gamma})
+\theta^{\mu\beta}\theta^{\gamma\nu}F^{\delta\alpha}F_{\delta\gamma}
F_{\alpha\beta}\right\}  \nonumber \\
& - & \frac 12 \left\{(\theta F)(\theta^{\gamma\beta}F^{\mu}_{~\gamma}
F^{\nu}_{~\beta}+\theta^{\nu\beta}F^{\mu\gamma}F_{\gamma\beta}-\theta^{\mu\beta}
F^{\nu\gamma}F_{\gamma\beta})+\theta^{\gamma\beta}\theta^{\mu\nu}
F^{\alpha\delta}F_{\delta\gamma}F_{\alpha\beta}\right\} \nonumber \\
& + & \frac 18 \,(\theta F)\left\{\theta^{\mu\nu}F^2+(\theta F)F^{\mu\nu}\right\}
 \nonumber \\
& + & \frac 14 \left(\theta^{\alpha\mu}\theta^{\nu\delta}F_{\alpha\delta}F^2+
\theta^{\alpha\beta}\theta^{\gamma\delta}F_{\alpha\delta}F_{\beta\gamma}
F^{\mu\nu}\right) \label{second}
\end{eqnarray}
This rather involved formula, when re-expressed in terms of the classical 
fields gives us the second order terms in $\theta$ to be added to the above: 
\begin{eqnarray}
\Db^{(2)} & = & \left[(\epsb\cdot\Eb-\betb\cdot\Bb)^2-\epsb^2\,\Bb^2+(\epsb\cdot\Bb)
^2+(\epsb\cdot\betb)(\Eb\cdot\Bb)\right]\,\Eb \nonumber \\
&  & +\,[(\betb\cdot\Eb)(\epsb\cdot\Eb-\betb\cdot\Bb)-(\epsb\cdot\Bb)
(\betb\cdot\Bb)+\betb^2(\Eb\cdot\Bb)+\frac{1}{2}\,(\epsb\cdot\betb)(\Eb^2-\Bb)^2
]\,\Bb \nonumber \\
&  & +\,\left[(\epsb\cdot\Eb-\betb\cdot\Bb)\Eb^2+(\betb\cdot\Eb)(\Eb\cdot\Bb)
+(\betb\cdot\Bb)\Bb^2\right]\,\epsb \nonumber \\
&  & +\,(\epsb\cdot\Eb-\betb\cdot\Bb)(\Eb\cdot\Bb)\,\betb+[\Eb\cdot
(\epsb\times\Bb)]\,\betb\times\Eb 
\end{eqnarray}
while for the magnetic induction we get:
\begin{eqnarray}
\Hb^{(2)} & = & [\epsb^2(\Eb\cdot\Bb)-(\epsb\cdot\Bb)(\epsb\cdot\Eb
-\betb\cdot\Bb)-(\epsb\cdot\Eb)(\betb\cdot\Eb)-\frac{1}{2}\,(\epsb\cdot\betb)
(\Eb^2-\Bb)^2]\,\Eb \nonumber \\
&  & +\,\left[(\epsb\cdot\Eb-\betb\cdot\Bb)^2-\betb^2\Eb^2+(\betb\cdot\Eb)^2+
(\epsb\cdot\betb)(\Eb\cdot\Bb)\right]\,\Bb \nonumber \\
&  & -\,(\epsb\cdot\Eb-\betb\cdot\Bb)(\Eb\cdot\Bb)\,\epsb
+[\Bb\cdot(\betb\times\Eb)]\,\epsb\times\Bb \nonumber \\
&  & +\,\left[(\epsb\cdot\Eb)\Eb^2+(\epsb\cdot\Bb)(\Eb\cdot\Bb)-
(\epsb\cdot\Eb-\betb\cdot\Bb)\Bb^2\right]\,\betb
\end{eqnarray}
We end this section observing that applying an Electric-magnetic duality 
directly on the classical fields and
reversely on the noncommuting parameter in this way:
\begin{equation}
\left\{\begin{array}{rcr}
\Eb & \rightarrow & -\Bb \\
\Bb & \rightarrow & \Eb 
\end{array}\right. ~~~~~~
\left\{\begin{array}{rcr}
\epsb & \rightarrow & \betb \\
\betb & \rightarrow & -\epsb
\end{array}\right.
\end{equation}
induces, up to second order, an "Electric-magnetic duality" on the NC 
fields (\ref{NC}):
\begin{equation}
\left\{\begin{array}{rcr}
\Db & \rightarrow & -\Hb \\
\Hb & \rightarrow & \Db 
\end{array}\right.
\end{equation}
At present, the meaning of this symmetry is unclear and we suspect it remains 
true to all orders in the perturbative $\theta$ expansion.

\newsection{Exact solutions and an iterative method}

We seek solutions to the equations of motion:
\begin{equation} \label{EQ}
\left\{
\begin{array}{rcl}
\partial_{[\nu} F_{\mu\rho]} & = & 0 \\
&  &  \\
\partial_\nu\tilde{F}^{\mu\nu} & = & 0 
\end{array}\right.
\end{equation}
where:
\begin{equation}  \label{sum1}   
\tilde{F}=F+\tilde{F}^{(1)}+\tilde{F}^{(2)}+\cdots
\end{equation}
with the structure $\tilde{F}^{(n)}\equiv\theta^n\,F^{n+2}$ already evident for 
example in eq. (\ref{first}) and (\ref{second}). \par
The most natural thing to suppose is that also a solution should be written 
as a sum:
\begin{equation}  \label{sum2}
F:=F^{(0)}+F^{(1)}+F^{(2)}+\cdots
\end{equation}
with pieces $F^{(k)}$ now understood to be corrections to a solution $F^{(0)}$ 
to the classical Maxwell Equations i.e. $\partial_\nu F^{(0)\mu\nu}=0$ plus the
Bianchi identities. 
We will briefly state this as $\partial F^{(0)}=0$. 
Furthermore, let $|_k$ be the operation of keeping, in a generic 
expression, all terms up to a given order $k$ in $\theta$, neglecting the others. 
Then extracting $k$-th order from (\ref{sum1}) terms like this:
\begin{equation}
F|_k=\sum_{i=0}^kF^{(i)}
\end{equation}
will be present. Accounting for that, hypotesis (\ref{sum2}) and the 
structure (\ref{struct}) we get:
\begin{equation}  \label{itera}
\tilde{F}|_k=F|_k+\theta\,(FF)|_{k-1}+\cdots+\underbrace{\theta\cdots\theta}
_k\,(\underbrace{F\cdots F}_{k+1})|_0
\end{equation}
Our purpose is to write down a recursive method of solving the noncommutative
Maxwell equation $\partial\tilde{F}=0$ having a classical solution $F^{(0)}$.
This is realized order by order noting that $(\partial\tilde{F})|_k=\partial
\tilde{F}|_k$. Then taking first order into the recursive relation (\ref{itera}) 
we have:
\begin{equation}
\partial\tilde{F}|_1=\partial(F^{(0)}+F^{(1)}+\theta(FF)|_0)=0
\end{equation}
Now, being $F^{(0)}$ a classical solution, we are led to solve the equation:
\begin{equation}  \label{uno}
\partial F^{(1)}=-\partial(\theta\,F^{(0)}F^{(0)})
\end{equation}
In exactly the same way, solutions correct up to second order come from:
\begin{equation}
\partial F^{(2)}=-\partial\left[\theta\,(F^{(0)}F^{(1)}+F^{(1)}F^{(0)})+
\theta\theta\,F^{(0)}F^{(0)}F^{(0)}\right]
\end{equation}
Generally, obtaining the $k$-th term in the expansion (\ref{sum2}) always 
reduces to solving an equation of the form:
\begin{equation}
\partial_\nu F^{(k)\mu\nu}=J^{\mu}[F^{(1)},\cdots ,F^{(k-1)}] \label{iter}
\end{equation}
where the right member $J^{\mu}$ only involves all the $k-1$ solutions 
computed in the previous steps. 
Now, deciding that each two form $F^{(k)}$ comes from a potential $A^{(k)}$
satisfying the Lorentz gauge constraint\footnote{It is easy to show that 
this can always be done order by order} $\partial_\nu A^{(k)\nu} =0$ then 
Eq.(\ref{iter})
becomes:
\eqn
-\square A^\mu =J^\mu
\feqn
This is immediately solved employing the Lienard-Wickert potentials. Then in
principle we have got an authomatic tool capable of solving the equations of 
motion in full. \par
Let us focus, for example, on the single plane wave solution:
\begin{equation}
A_\mu=\zeta_\mu e^{i\,k\cdot x}
\end{equation}
with $k_\mu k^\mu=\zeta_\mu k^\mu=0$ in the Lorentz gauge 
$\partial_\nu A^\nu=0$. We have:
\begin{equation}
F^{(0)}_{\mu\nu}=i\,(k_\mu\zeta_\nu-k_\nu\zeta_\mu) e^{i\,k\cdot x}
\end{equation}
This is a particular case because we will now show that it is an {\bf exact}
solution of eq.(\ref{EQ}).
\begin{lem}   \label{lemma}
Given an antisymmetric matrix $\theta^{\mu \nu}$, a null vector $k^\alpha$ and
a family of vectors $\{ \zeta^\beta_{(i)} \}_{i\in I}$ orthogonal to
$k^\alpha$, then any combination
of n copies of $\theta^{\mu \nu}$, (n+1) vectors of the given family and
(n+2) copies of $k^\alpha$ in which all indices but one are saturated, vanish.
\end{lem}
{\em Proof.} Try to build a nonvanishing combination. In so doing, you cannot
saturate the $k$ vectors with the $\zeta$ vectors due to ortogonality. Neither
you can saturate two of them with one $\theta$ matrix due to its antisimmetry.
You are obliged to use one only $k$ vector for each matrix, spending $n$ of them.
Of the two remaining, one can be chosen as the free index but the other
must necessarily be saturated with one of the $\zeta$ vectors or one of the
$\theta$ matrices giving a vanishing result. $\qquad$ QED.
\begin{prop}
Monochromatic plane waves solve the field equations (\ref{EQ}) to every 
order in $\theta$.
\end{prop}
{\em Proof.} Let us write the general monochromatic plane wave as:
\eqn
A_\mu =\Phi_\mu (K \cdot x)
\feqn
with $K^2 =0$ and $K^\mu \Phi^\prime_\mu  (K \cdot x) =0$
so that
\eqn
F_{\mu \nu} =K_\mu \Phi^\prime_\nu  (K \cdot x) -K_\nu \Phi^\prime_\mu
(K \cdot x)
\feqn
 \\
Let $\tilde {F}^{(n) \mu \nu}$ be the term of order $n$ in $\theta$; then
$\partial_\mu \tilde {F}^{(n) \mu \nu}$ is the sum of terms obtained by
contraction of $n$ copies of $\theta^{\mu \nu}$, $n$ copies of
$\Phi^\prime_\alpha$, one copy of $\Phi^{\prime \prime}_\beta$ and $n+2$
copies of $K_\gamma$. From the Lemma it follows that
$\partial_\mu \tilde {F}^{(n) \mu \nu}=0 \ . \qquad$ QED. \\
The previous property of monochromatic plane waves holds for any
lagrangian having the assumed polynomial structure,
independently of the fact that it has been derived from a NC theory using
the SW map.

\subsection{Plane wave superposition}

While single plane waves turn out to be exact solutions of the field equations
this is no longer valid even for a simple superposition like this:
\begin{equation}
A_\mu:=\zeta_\mu e^{i\,k\cdot x}+\zeta'_\mu e^{i\,k'\cdot x}
\end{equation}
corresponding to the classical solution (by linearity):
\begin{equation}
F^{(0)}_{\mu\nu}:=i\,(k_\mu\zeta_\nu-k_\nu\zeta_\mu) e^{i\,k\cdot x}+
i\,({k'}_\mu{\zeta'}_\nu-{k'}_\nu{\zeta'}_\mu) e^{i\,k'\cdot x}
\end{equation}
To find out its first order correction in the NC framework we must solve 
(\ref{uno}) yielding:
\begin{eqnarray}
\partial_\nu F^{(1)\mu\nu} & = & i\left\{
\left[(k k')_\theta (\zeta\zeta')-(\zeta\zeta')_\theta (k k')\right]
(k^\mu -{k'}^\mu
)
\right. \cr
&  & ~~~-
\left[(k\zeta')_\theta(\zeta k')-(\zeta k')_\theta(k\zeta')\right](k^\mu+
{k'}^\mu
)\cr
&  & ~~~+2\,
\left[(k\zeta')_\theta(k k')-(k k')_\theta(k\zeta')\right]\zeta^\mu \cr
&  & ~~~-2\,
\left[(\zeta k')_\theta(k k')-(k k')_\theta(\zeta k')\right]{\zeta'}^\mu
\left.\right\}e^{i\,(k+k')\cdot x}  \label{finire}
\end{eqnarray}
where the following anti-symmetric inner product has been defined: $(vw)_\theta
:=v^\mu\,\theta_{\mu\nu}\,w^\nu$.\par\noindent
This equation can be solved assuming
\begin{equation}
F^{(1)}_{\mu\nu}=\partial_\mu A^{(1)}_{\nu}-\partial_\nu A^{(1)}_{\mu}
\end{equation}
and $A^{(1)}$ still satisfying an extended Lorentz gauge constraint
$\partial_\nu A^{(1)\nu}=0$. \\
Infact, rewriting eq. (\ref{finire}) in the form 
\eqn
-\square A^{(1)}_\mu=iJ_\mu e^{i\,(k+k')\cdot x} 
\feqn
we realize that $J$ is transverse to $k+k'$ as can be easily proved using
the defining relations:
\eqn
k^2={k'}^2=0 \qquad k\cdot\zeta=k'\cdot\zeta'=0
\feqn
This means that if we put abruptely,
\eqn
A^{(1)}_\mu=\frac {iJ_\mu}{2k\cdot k'}\,e^{i\,(k+k')\cdot x} \label{sup}
\feqn
this solves eq.(\ref{finire}) being also compatible with the extended 
Lorentz gauge. \par
Note that this corrected version of the superposition law
could be used to reveal a refraction effect suffered by a ray of light
in passing from an empty region to one in which a background static 
and uniform magnetic or electric field is present \cite{JK1,JK2}}.
The incoming and reflected rays propagating in the empty region,
should be described by a superposition of waves agreeing with the
refracted one in the transition region.

\subsection{The Coulomb Law}
\begin{figure}[h]
\centering
\epsfig{file=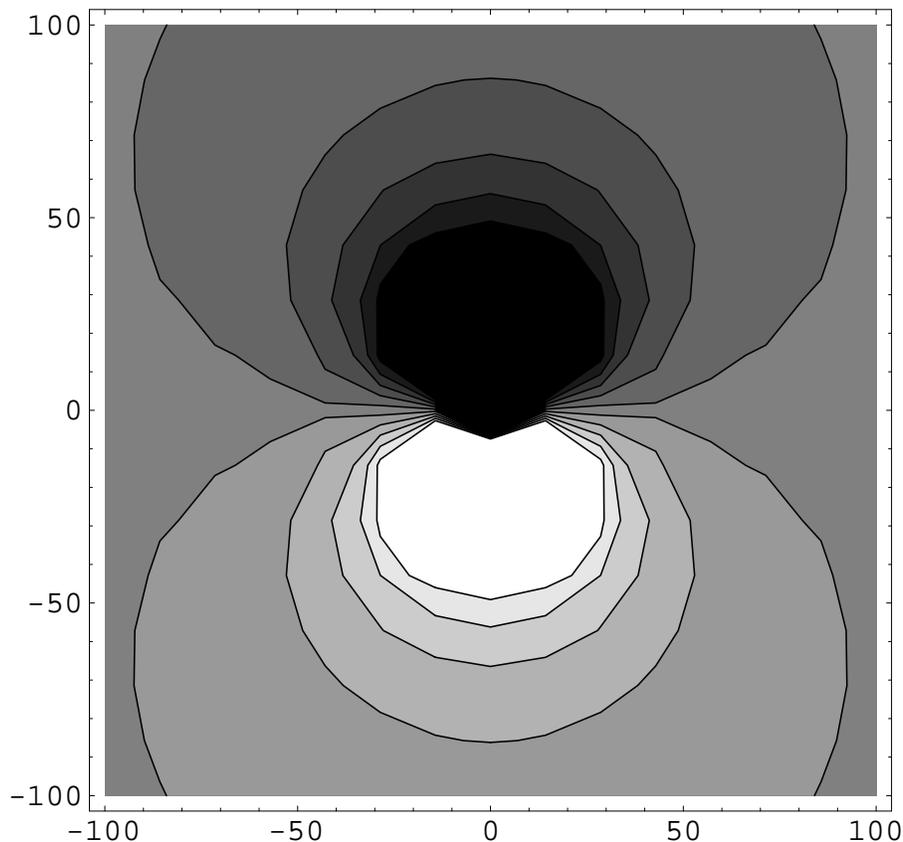,height=12cm,width=12cm,
        bbllx=0pt,bblly=0pt,bburx=290pt,bbury=290pt}
\caption{The correction to the Coulomb law due to $A^{(1)}$. The figure
  shows the equipotential levels in the $(x,z)$ plane, assuming that 
$\vec\epsilon$ is oriented along the $z$-axis.
We have chosen $\epsilon=1,~r_0=1$. The length unit is $r_0$.
 }
\label{figA1}
\end{figure}

All NC theories are characterized by a parameter $\theta$ which
defines a natural scale of length.
From a dimensional analysis,
the corrections to the Coulomb law are of order $\frac 1{r^4}$ but 
a complete power series expansion in $\frac \theta{r^2}$ is expected so that
if $L$ is its convergence ratio (plausibely finite), then non-perturbative 
contributions should become relevant in the region $r < \sqrt
{\theta}{L}$ where the perturbative description fails.
We can make a sensible study of the NC corrections to the Coulomb law,
considering the potential generated by a charged conducting sphere of
radius $r_0$.\\
At $0$-th order the classical potential is
\eqn
A^{(0)}=
\begin{cases}    
-\frac er \,dt & r>r_0 \\ 
-\frac {e}{r_0} \,dt & r\leq r_0 \\ 
\end{cases}
\label{0coulomb}
\feqn
so that ($\hat x^i$ is the radial versor)
\eqn
F^{(0)}_{0i} =-\frac e{r^2} \,\hat x^i~\theta(r-r_0)
\label{Abou-Zeid:2001up}
\feqn
At first order:
\eqn
\partial_\nu F^{(1) \mu \nu} =-\partial_{\nu}
G^{\mu \nu}  \label{iteracoulomb}
\feqn
with
\eqn
G^{\mu \nu} := -\frac 12 (\theta F) F^{\mu \nu} -\frac 14 \,\theta^{\mu \nu}
(FF) +\theta_{\alpha \beta} F^{\mu \alpha} F^{\nu \beta} +
(\theta^{\mu \beta} F^{\alpha \nu} -\theta^{\nu \beta} F^{\alpha \mu}
)F_{\alpha \beta}
\feqn
A direct computation gives for the tensor $G$:
\eqn
& & G^{0i} =-\frac {e^2}{r^4} \left[ (\epsilon \cdot \hat x ) \hat x^i
+\frac 12 \,\epsilon^i \right]~
\theta(r-r_0)  \\
& & G^{ij} =\frac {e^2}{r^4} \left[(\hat x^i \theta^{jk} -\hat x^j \theta^{ik} )
\hat {x}_k + \frac 12 \,\theta^{ij}  \right]~\theta(r-r_0)
\feqn
so that
\eqn
\partial_\mu F^{(1) \mu \nu} =J^\nu  \label{risolvere}
\feqn
with
\eqn
J= \frac {e^2}{r^6}\, (4\vec \epsilon \cdot \vec r ;\vec r \wedge \vec
\beta ) ~\theta(r-r_0) ~+~
\frac{e^2}{r_0^5} (3/2~ \vec \epsilon \cdot \vec r; -1/2~ \vec r \wedge
\vec\beta  ) ~\delta(r-r_0)
\label{current}
\feqn
We have solved numerically the equation \ref{risolvere} 
and we show in Figure \ref{figA1} the equipotential level of the
zeroth component of $A^{(1)}$ which is symmetrical under rotations
about the direction of $\vec\epsilon$.
The other components have a similar angular behaviour and their precise values
depend on the direction and magnitude of $\beta$.\\
The corrections in Figure \ref{figA1} give the modification of the
Coulomb law in the case of a charged sphere.
This interpretation would be well defined with $r_0\gg\sqrt\theta$, being
the perturbative solution valid almost everywhere, even inside the
conducting sphere.
But in the specific example considered in Figure \ref{figA1} the sphere
coincides with the excluded region, where the perturbative approach
fails. This case suggests a different interpretation, as the NC
correction to the Coulomb potential of a point-charge.
In fact in NC theories it is intuitive to replace pointlike with
extended object, whose typical length is $\sqrt\theta$.\\
The corrections to the potential violate the Gauss law and the
spherical symmetry of the classical solution.
As a consequence, we observe in the case of a conducting macroscopic sphere, 
that the potential inside the conductor is not constant.
This remark suggests a way to test NC electrodynamics effects.
\begin{figure}[h]
\centering
\epsfig{file=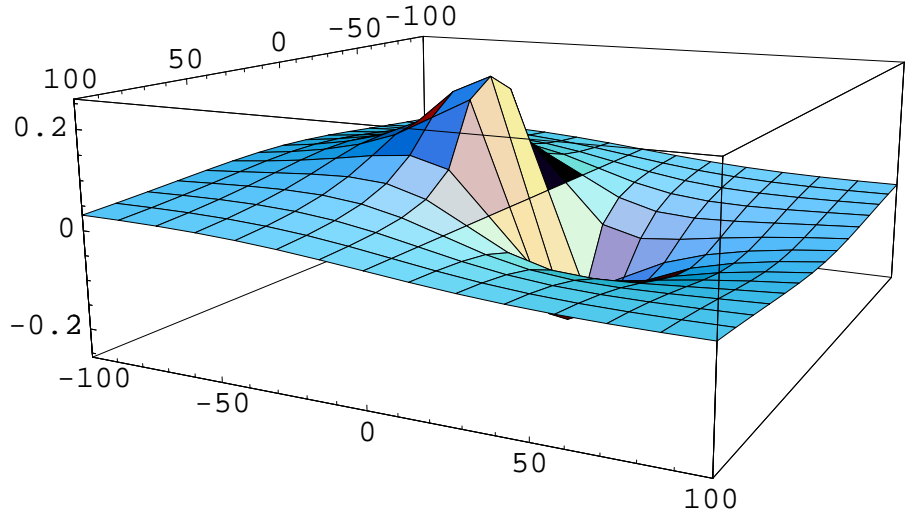,height=12cm,width=12cm,
        bbllx=0pt,bblly=0pt,bburx=290pt,bbury=290pt}
\caption{The ratio $A^{(1)}/A^{(0)}$ which shows the effect of the 
NC corrections to the Coulomb law, in the $(x,z)$ plane.
We have chosen $\vec\epsilon$ oriented along the $z$-axis,
$\epsilon=1, \ e=1$ and $r_0=1$. The length unit is $r_0$.
 }
\label{ratio}
\end{figure}
In Figure \ref{ratio} we show the relative contribution of the
corrections $A^{(1)}$ to the classical Coulomb potential $A^{(0)}$.
The size of the corrections is already relevant (e.g. greater than 10
percent) at a length scale bigger by more than one order of magnitude
w.r.t. to the one determined by the NC parameter.

\section{Conclusions}

In this work we have formulated an explicit perturbative realization
of NC electrodynamics,
which turns out to be causal and Lorentz invariant.
The basic steps to obtain this result have been:
the use of the SW map and
a rearrangement of the action aimed to render every term explicitly
gauge invariant, by use of careful integration by parts.
The resulting expressions do not contain time derivatives of order
higher than two, yielding authomatically a causal theory.
This latter property is obtained without imposing any constraint on
the NC parameter $\theta$, which can be chosen in full generality as a Lorentz
tensor, leading to a Lorentz covariant theory.\\
We have studied the general structure of the Lagrangian, to all orders
in the perturbative expansion.
We have shown that the monochromatic plane wave is solution of the
equations of motion to first \cite{JK2} and even to all orders.\\
We developed an iterative method to solve the equations of motion.
In particular we applied this method to study the corrections to the
superposition law of plane waves and to the electrostatic potential of
a spherically symmetric charge distribution. 
The most relevant qualitative feature of the NC corrections that we
calculated is that they have a peculiar signature which makes them, at
least in priciple, distinguishable from the classical corresponding
effects.
A possible test of the superposition law could be done by studying the
reflection and rifraction of light on a magnetic field,
using for instance the experimental setting described
in \cite{Zv}.
Furthermore, the deviations from the Coulomb law could be evidenced by 
measuring the charge distribution on the surface and the electric
field  inside an empty conducting sphere.
\vfill
{\bf Acknowledgements\\}
We are grateful to D. Klemm for his participation in the initial stage
of this work. This work was partially supported by INFN, MURST and by
the European Commision RTN program HPRN-CT-2000-00131, in which
the authors are associated to the University of Torino.

\end{document}